# Distributed Spiral Optimization in Wireless Sensor Networks without Fusion Centers

Zheng SUN

Key Laboratory of Information Processing and Intelligent Technology

Beijing University of Posts and Telecommunications, Beijing, China

zhengs.bupt@gmail.com

*Abstract*：A distributed spiral algorithm for distributed optimization in WSN is proposed. By forming a spiral-shape message passing scheme among clusters, without loss of estimation accuracy and convergence speed, the algorithm is proved to converge with a lower total transport cost than the distributed in-cluster algorithm.

*Introduction*: The most straight forward way to solve optimization problems in wireless sensor networks (WSN) is by centralized processing, where all nodes send observations to one central point that solves the problem using algorithms introduced by [1]. However, this approach is time-consuming, and is not energy-efficient. Recently, alternative decentralized approaches have been proposed, which adopt various distributed optimization algorithms. A brilliant solution has been originally introduced in [2] and further developed in [3]. Furthermore, the decentralized approaches can also be divided into in-network and in-cluster schemes. The former considers the network as a whole, and utilizes peer-to-peer communication to solve the problem [4]. And the latter, which is introduced in [5], is a hybrid form of both the centralized scheme and the in-network scheme. It can achieve higher estimation accuracy, lower transport cost, and lower latency than the distributed in-network schemes. But the drawback is the involvement of fusion centers and thus the low robustness of the system against fusion center failure and the high sensor-to-fusion-center transport costs. In this paper, we propose a distributed spiral algorithm, which achieves lower transport costs without fusion centers.

*Distributed Spiral Algorithm*: Let us consider a convex optimization problem with the form

$$\begin{aligned}\min_{\theta} \quad & \sum_{i=1}^{n} f_i(x_i, \theta) \\ \text{subject to} \quad & \theta \in \Theta\end{aligned} \quad (1)$$

where $f_i(x_i, \theta)$ is a cost function associated with node $i$, $x_i$ is a vector of "private" observations local to node $i$, and $\theta$ is a global decision variable. Assume that $f_i$ are convex functions and $\Theta$ is a convex set with non-empty interior. Consider a WSN with $n_C$ clusters and $n_S$ sensors per cluster. We use $i$ to index sensors, $j$ to clusters, and $k$ to the iteration numbers. Let $i=0$ indicate cluster head. Denote $f_{i,j}$ and $\varphi_{i,j,k}$ as the local cost function and the estimate of $\theta$ at node $i$ in cluster $j$ at iteration $k$, respectively.

In iteration $k$, the distributed spiral algorithm proceeds as follows:

1) Cluster heads initialize $\varphi_{0,j,k} = \theta_{j,k}$.

2) Incremental update is conducted in parallel in all clusters, and in each cluster, the update: $\varphi_{i,j,k} = \varphi_{i-1,j,k} + \alpha_k g_{i,j,k}$, are conducted through a path that traverse all the nodes in each cluster, where $\alpha_k$ is the step-size and $g_{i,j,k}$ is a subgradient of $f_{i,j}$ with $\varphi_{i-1,j,k}$ and $x_{i,j}$. (When $\varphi_{i-1,j,k} \notin \Theta$, a projection is needed.)

3) Rather than passing estimates to the fusion center as described in [5], every cluster head passes the last in-cluster estimate $\varphi_{n_S,j,k}$ to the former cluster, which computes the next estimate by $\theta_{j,k+1} = (\varphi_{n_S,j,k} + \varphi_{n_S,(j+1)\bmod n_C,k})/2$. Then repeat.

Note that in Step 1 and 2 of every iteration, sensor nodes conduct the same intra-cluster update as that of the distributed in-cluster algorithm, but at the end of each iteration, every cluster head transmits the last estimate to the cluster head of the former cluster in parallel, thus form a spiral-shape update. With the change of step 3, the algorithm needs only inter-sensor and inter-cluster transmissions, and therefore eliminates fusion centers.

The distributed in-cluster algorithm and the distributed spiral algorithm are depicted in Figure 1.

In what follows, the convergence of the distributed spiral algorithm under fixed step-sizes is proved. Under diminishing step-sizes, the algorithm converges to the global optimal value $f(\theta^*)$. Due to the space limit, the latter proof is omitted here. Assume $\forall i, j$, $\|g_{i,j,k}\|$ is upper bounded by an existing non-negative scalar $C_{i,j}$. In the proofs below, for clarity, we denote $\sum_{i=1}^{n_S} C_{i,j}$ as $C_j$, and $\sum_{j=1}^{n_C} \sum_{i=1}^{n_S} C_{i,j}$ as $C$.

*Lemma 1*: Denote $\sum_{j=1}^{n_C} \left(\sum_{i=1}^{n_S} C_{i,j}\right)^2$ as $\hat{C}$, we have

$$\sum_{j=1}^{n_C} \|\theta_{j,k+1} - y\|^2 \leq \sum_{j=1}^{n_C} \|\theta_{j,k} - y\|^2 - 2\alpha_k \sum_{j=1}^{n_C} \left(f_j(\theta_{j,k}) - f_j(y)\right) + \alpha_k^2 \hat{C}, \quad (2)$$

A proof is presented in the Appendix. Lemma 1 guarantees that the average distance between $\{\theta_{j,k}\}_{k=1}^{\infty}$ and $y$ gets smaller provided that $\alpha_k < 2\sum_{j=1}^{n_C}\left(f_j(\theta_{j,k}) - f_j(y)\right)\big/\hat{C}$.

*Theorem 1*: $\forall j$, let $\{\theta_{j,k}\}_{k=1}^{\infty}$ be a sequence generated by the distributed spiral algorithm. Then, for a fixed step-size, $\alpha_k = \alpha$, we have

$$\liminf_{k \to \infty} \sum_{j=1}^{n_C} f_j(\theta_{j,k}) \leq f(\theta^*) + \frac{1}{2}\alpha \sum_{j=1}^{n_C} C_j^2, \quad (3)$$

where $f(\theta^*) = \inf_{\theta \in \Theta} f(\theta)$.

A proof is in the Appendix. Theorem 1 guarantees the convergence of the distributed spiral algorithm [6].

*Performance Analysis*: In [5], the authors have proved that the distributed in-cluster algorithm achieves higher energy efficiency, lower latency, and higher estimation accuracy than both the centralized and the distributed in-network algorithms. So in this section, we focus on analyze the performance difference between the distributed spiral algorithm and the distributed in-cluster algorithm. For energy efficiency, consider a WSN consisting of $n$ nodes and each node collects $m$ observations. The nodes are uniformly distributed within a unit square meter. Assume the network is divided into $n_C$ clusters with $n_S = n/n_C$ nodes per cluster. As same as in [5], we adopt bit-meter as the metric to measure the transport cost in the transmission of data. The distributed in-cluster algorithm requires a sensor-to-sensor transport cost of $O(n)$ bits to be transmitted over an average distance of $O(1/\sqrt{n})$ meters, and $O(n_C)$ bits to be transmitted over an average distance of $O(1)$ meters, which accounts for the cluster-head-to-fusion-center transport cost. Thus, the total transport cost is $O(\sqrt{n}+n_C)$. In the distributed spiral algorithm, the sensor-to-sensor transport cost is the same as the distributed in-cluster algorithm, but the cluster-head-to-fusion-center transmission is substituted by an inter-cluster transmission, which has an average distance of $O(1/\sqrt{n_C})$ meters. Thus, the total transport cost is reduced to $O(\sqrt{n}+\sqrt{n_C})$. For latency, the latency of the in-cluster algorithm is adjustable and determined by the size of the cluster. Since the number of nodes within clusters is the same, the distributed spiral algorithm has the same latency as the distributed in-cluster algorithm. For estimation accuracy, Theorem 1 shows that the distributed spiral algorithm achieves exactly the same estimation accuracy as the distributed in-cluster algorithm in [5].

*Simulations*: In a WSN with 64 sensors uniformly distributed, the nodes are divided into 8 clusters. Every node takes 10 observations, and all observations are i.i.d. We conduct the same least squares estimation as described in Section VI of [5]. Fig. 2 shows simulation results with a diminishing step-size of $\alpha_k = 0.007/k$ and a fixed step-size of $\alpha_k = 0.007$ with "-d" and "-f", respectively. Both results show that the distributed spiral algorithm has almost exactly the same estimation accuracy and convergence speed as the distributed in-cluster algorithm.

*Conclusions*: In this paper, we have proposed a distributed spiral algorithm, which does not need the involvement of fusion centers to perform distributed optimization in WSN. By forming a spiral-shape message passing scheme among clusters, the total transport cost has been reduced without loss of estimation accuracy and convergence speed.

*Acknowledgement*: This work is sponsored by Project 60772108 supported by National Natural Science Foundation of China and National Basic Research Program of China (973 Program), No. 2007CB310604.

**Appendix**

*Proof of Lemma 1*: Using the result from [6]. Let $\{\varphi_{i,j,k}\}, \forall i, j$, be the sequence of subiterations, then for all $y \in \Theta$ and for $k \geq 0$, we have $\|\varphi_{i,j,k} - y\|^2 \leq \|\varphi_{i-1,j,k} - y\|^2 - 2\alpha_k \left( f_{i,j}(\varphi_{i-1,j,k}) - f_{i,j}(y) \right) + \alpha_k^2 C_{i,j}^2$. By summing all the inequalities over all $i$ and $j$, we have

$$\|\varphi_{n_S, j, k} - y\|^2 \leq \|\varphi_{0, j, k} - y\|^2 - 2\alpha_k \sum_{i=1}^{n_S} \left( f_{i,j}(\varphi_{i,j,k}) - f_{i,j}(y) \right) + \alpha_k^2 \sum_{i=1}^{n_S} C_{i,j}^2$$
$$\leq \|\varphi_{0,j,k} - y\|^2 - 2\alpha_k \left( f_j(\varphi_{0,j,k}) - f_j(y) \right) - 2\alpha_k \sum_{i=1}^{n_S} \left( f_{i,j}(\varphi_{i,j,k}) - f_{i,j}(\varphi_{0,j,k}) \right) + \alpha_k^2 \sum_{i=1}^{n_S} C_{i,j}^2 \quad . \quad (4)$$

Since $f_{i,j}(\varphi_{0,j,k}) - f_{i,j}(\varphi_{i,j,k}) \leq C_{i,j} \|\varphi_{0,j,k} - \varphi_{i,j,k}\|$ and $\|\varphi_{i,j,k} - \varphi_{0,j,k}\| \leq \alpha_k \sum_{p=1}^{i} C_{p,j}$, equation (4) becomes

$$\|\varphi_{n_S, j, k} - y\|^2 \leq \|\varphi_{0,j,k} - y\|^2 - 2\alpha_k \left( f_j(\varphi_{0,j,k}) - f_j(y) \right) + 2\alpha_k^2 \sum_{i=1}^{n_S} C_{i,j} \sum_{p=1}^{i} C_{p,j} + \alpha_k^2 \sum_{i=1}^{n_S} C_{i,j}^2$$
$$\leq \|\varphi_{0,j,k} - y\|^2 - 2\alpha_k \left( f_j(\varphi_{0,j,k}) - f_j(y) \right) + \alpha_k^2 C_j^2 \quad , \quad (5)$$

Because $\theta_{j,k+1} = \frac{1}{2}\left( \varphi_{n_S,j,k} + \varphi_{n_S,(j+1) \bmod n_C, k} \right)$, let us denote $(j+1) \bmod n_C$ as $j'$, so

$$\|\theta_{j,k+1} - y\|^2 \leq \frac{1}{2}\left( \|\varphi_{n_S, j, k} - y\|^2 + \|\varphi_{n_S, j', k} - y\|^2 \right). \quad (6)$$

Put (5) into (6) and notice that $\theta_{j,k} = \varphi_{0,j,k}$. Then by summing all the inequalities in (6) over all $j$, we have the relation stated in (2).

*Proof of Theorem 1*: Proof is by contradiction. If Theorem 1 does not hold, then there must at least exist an $\varepsilon > 0$ such that

$$\liminf_{k \to \infty} \sum_{j=1}^{n_C} f_j(\theta_{j,k}) > f(\theta^*) + \frac{1}{2}\alpha \sum_{j=1}^{n_C} C_j^2 + 2\varepsilon. \qquad (7)$$

Let $\hat{y} \in \Theta$ be such that

$$\liminf_{k \to \infty} \sum_{j=1}^{n_C} f_j(\theta_{j,k}) \geq f(\hat{y}) + \frac{1}{2}\alpha \sum_{j=1}^{n_C} C_j^2 + 2\varepsilon, \qquad (8)$$

and let $k_0$ be large enough so that for all $k \geq k_0 > K$ we have

$$\sum_{j=1}^{n_C} f_j(\theta_{j,k}) \geq \liminf_{k \to \infty} \sum_{j=1}^{n_C} f_j(\theta_{j,k}) - \varepsilon. \qquad (9)$$

By adding the preceding two inequalities, we obtain for all $k \geq k_0$,

$$\sum_{j=1}^{n_C} f_j(\theta_{j,k}) - f(\hat{y}) \geq \frac{1}{2}\alpha \sum_{j=1}^{n_C} C_j^2 + \varepsilon. \qquad (10)$$

Put (10) into (2) and by Lemma 1 for the case where $y = \hat{y}$ together with the above relation, we obtain for all $k \geq k_0$,

$$\sum_{j=1}^{n_C} \|\theta_{j,k+1} - \hat{y}\|^2 \leq \sum_{j=1}^{n_C} \|\theta_{j,k} - \hat{y}\|^2 - 2\alpha\varepsilon. \qquad (11)$$

Summing all inequalities over all $k \geq k_0$, we have

$$\sum_{j=1}^{n_C} \|\theta_{j,k+1} - \hat{y}\|^2 \leq \sum_{j=1}^{n_C} \|\theta_{j,k_0} - \hat{y}\|^2 - 2(k+1-k_0)\alpha\varepsilon, \qquad (12)$$

which cannot hold for sufficiently large $k$ – a contradiction.

**Author's Affiliations**

Zheng SUN (Key Laboratory of Information Processing and Intelligent Technology, Beijing University of Posts and Telecommunications, Beijing, China)

**Email Address**

zhengs.bupt@gmail.com

**Figure Captions**

Fig. 1   Illustration of distributed optimization algorithms. Left: Distributed in-cluster algorithm. Right: Distributed spiral algorithm.

Fig. 2   Simulation results of least squares estimation with diminishing and fixed step-sizes.

Figure 1

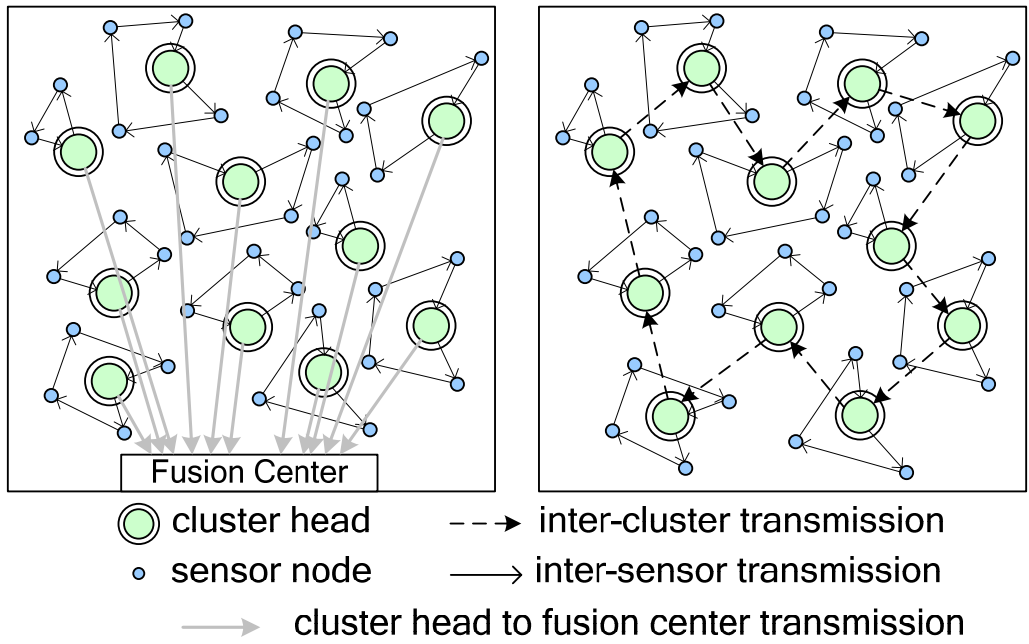

Figure 2

![Figure 2: Plot showing $\theta_j$ for different clusters versus Number of Iterations, comparing in-cluster and spiral methods for clusters #1, #4, #7 with d and f variants, converging to exact value around 14.5]